\newcommand{\apj} {{\it ApJ}}
\newcommand{\apjl} {{\it ApJL}}
\newcommand{\apjs} {{\it ApJS}}
\newcommand{\araa} {{\it ARA\&A}}
\newcommand{\mnras} {{\it MNRAS}}

\newcommand{\ea} {{et al.}}
\newcommand{\hii} {H{\sc ii}}
\newcommand{\kms}{{\rm ~km~s}^{-1}}
\newcommand{\Lb} {L_{\rm box}}

\newcommand{\Mmax} {M_{\rm max}}
\newcommand{\Ms}{M_{\rm s}}
\newcommand{\Msun}{{\rm M}_{\odot}}

\newcommand{\pcc} {{\rm cm}^{-3}}
\newcommand{\VS} {{V\'azquez-Semadeni}}

\documentclass{iau} 
\usepackage{graphicx}

\title[Cluster Assembly in Collapsing Clouds] 
{Cluster Assembly in Hierarchically Collapsing Clouds}

\author[V\'azquez-Semadeni et al.]   
{Enrique V\'azquez-Semadeni$^1$, Alejandro
Gonz\'alez-Samaniego$^2$, Manuel Zamora-Avil\'es$^1$ 
 \and Pedro Col\'in$^1$}

\affiliation{$^1$Instituto de Radioastronom\'ia y Astrof\'isica, UNAM \\
Antigua Carretera a P\'atzcuaro \# 8701, Morelia, Michoac\'an, 58088, M\'exico
\\ email: {\tt e.vazquez,m.zamora,p.colin@crya.unam.mx} \\[\affilskip]
$^2$Instituto de Astronom\'ia, UNAM,\\
Circuito Exterior, S/N,
M\'exico, D.F., 04510, M\'exico
\\email: {\tt ags@astro.unam.mx}}

\pubyear{2015}
\volume{xxx}  
\jname{IAU Symposium 316: Clusters}
\editors{A.C. Editor, B.D. Editor \& C.E. Editor, eds.}
\begin{document}

\maketitle

\begin{abstract}
  We discuss the mechanism of cluster formation in hierarchically
  collapsing molecular clouds. Recent evidence, both observational and
  numerical, suggests that molecular clouds (MCs) may be undergoing
  global, hierarchical gravitational collapse. The ``hierarchical''
  regime consists of small-scale collapses within larger-scale ones. The
  latter implies that the star formation rate increases systematically
  during the early stages of evolution, and occurs via filamentary flows
  onto ``hubs'' of higher density, mass, and velocity dispersion, and
  culminates a few Myr after than the small-scale collapses have started
  to form stars. In turn, the small-scale collapses occur in clumps
  embedded in the filaments, and are themselves falling into the larger
  potential well of the still-ongoing large-scale collapse. The stars
  formed in the early, small-scale collapses share the infall motion of
  their parent clumps towards the larger potential trough, so that the
  filaments feed both gaseous and stellar material to the hubs. This
  leads to the presence of older stars in a region where new protostars
  are still forming, to a scale-free or fractal structure of
  the clusters, in which each unit is composed of smaller-scale ones,
  and to the eventual merging of the subunits, explaining the
  observed structural features of open clusters.

  \keywords{Galaxies: star clusters, Gravitation, Hydrodynamics, ISM:
    clouds, Stars: formation}
\end{abstract}

\firstsection 

\section{Introduction}

It is presently accepted that most stars form in clusters, although the
details of the cluster-formation process, especially the origin of their
structural properties, remain a matter of active research (e.g.,
\cite[Lada \& Lada 2003]{LL03}). In particular, a
number of structural properties of the clusters have been uncovered that
still require an adequate theoretical understanding, such as: i) the
existence of a mass segregation in the clusters, with the most massive
stars lying closer to the cluster's center (\cite[Hillenbrand \&
Hartmann 1998] {HH98}); ii) the distribution of protostellar
separations, which appears to have no characteristic scale
(\cite[Bressert \ea\ 2010] {Bressert+10}); iii) the likely existence of
an age gradient in clusters, with the youngest stars being located in
the highest-density regions (\cite[Kuhn \ea\ 2015a] {Kuhn+15a}).

Numerical simulations have begun to offer some insight about these
properties. \cite[Kirk \ea\ (2014)] {Kirk+14} have concluded, from a
suite of simulations of self-gravitating, decaying isothermal
turbulence, that the most massive stars form in situ at the cluster
centers, rather than ``sinking'' there through dynamical interactions in
the cluster itself. However, they gave no physical explanation as to why
the most massive stars should form there. More recently, \cite[Kuhn \ea\
(2015b)]{Kuhn+15b} have suggested, by comparing multi-wavelength
observations of stellar clusters with numerical simulations, that
clusters form by mergers of ``subcluster'' structures, although again no
explanation of why such mergers should occur is provided. The presence
of subunits of somewhat different ages in the clusters has also been
pointed out by \cite[Rivera-G\'alvez et al.\ (2015)] {RG+15}.

A physical mechanism capable of providing a unifying scenario to
these properties is the one advanced by \cite[V{\'a}zquez-Semadeni et
al.\ (2009)] {VS+09; see also \G\'omez \& \VS\ 2014}. These authors
proposed that, if molecular clouds (MCs) are assembled by large-scale
colliding streams of warm, atomic gas, they rapidly condense into the
cold atomic phase, becoming Jeans-unstable and beginning to collapse
globally. Moreover, the collision of the
streams causes moderately supersonic turbulence (\cite[e.g., Koyama \&
Inutsuka 2002; Heitsch \ea\ 2005] {KI02, Heitsch+05}) in the cold gas,
which produces a spectrum of density fluctuations, where
large-amplitude, small-scale (LASS) fluctuations are superposed on
smaller-amplitude, larger-scale (SALS) ones (e.g., \cite[Kim \& Ryu 2005]
{KR05}). Since these density fluctuations are nonlinear, the LASS
fluctuations have shorter free-fall times than the SALS ones, therefore
completing their collapse earlier.  This process is therefore similar to
\cite[Hoyle's (1953)] {Hoyle53} fragmentation, except that the density
fluctuations are of turbulent origin and are therefore nonlinear. In what
follows, we will refer to this scenario as ``hierarchical gravitational
collapse'' (HGC).

This scenario also predicts that the star formation rate (SFR) in
molecular clouds (MCs) evolves over time, as a consequence of the
evolution of the clouds themselves as they go through global
gravitational collapse (\cite[Zamora-Avil\'es \ea\ 2012] {ZA+12};
\cite[Hartmann \ea\ 2012] {Hartmann+12}; \cite[Zamora-Avil\'es \& \VS\
2014] {ZV14}).

In this contribution, we describe results of a model on the evolution of
the SFR and of cluster assembly in a numerical simulation of MC
formation and evolution, emphasizing the role of HGC in the resulting
structure of the cluster.

\section{A model for the evolution of the SFR in molecular 
clouds} \label{sec:mod_SFR}

Several models for the SFR and the core mass function (CMF) have been
put forward in the last decade (e.g., \cite[Krumholz \& McKee 2005]
{KM05}; \cite[Hennebelle \& Chabrier 2011] {HC11}; \cite[Padoan \&
Nordlund 2011] {PN11}), all based on the notion that MCs are globally
supported by supersonic turbulence, and with the SFR given by the
collapse of local density fluctuations induced by the turbulence.

Operationally, the computation of the SFR in these models is performed
by using the probability density function (PDF) of the density
field---which is known to have a lognormal form for a
non-self-gravitating, isothermal turbulent gas (\cite[\VS\ 1994]
{VS94})---to compute the mass fraction at densities above a suitably
defined density threshold, and dividing it by a timescale characteristic
of its density range. Details of the similarities and differences
between these models have been given by \cite[Federrath \& Klessen
(2012)] {FK12}. Models of this kind are intrinsically {\it stationary},
as they assume that the MCs are in approximate virial equilibrium
between turbulence and gravity. In these models, the main controlling
parameters are those characterizing the turbulence -- the Mach number
$\Ms$, the {\it virial parameter} $\alpha$ (\cite[Bertoldi
\&McKee 1992] {BM92}), and perhaps the ratio of compressible to
solenoidal energy (\cite[Federrath \& Klessen 2012] {FK12}).

However, the basic premise of equilibrium in the stationary models is
inconsistent with the scenario of HGC, since in the latter, the cloud is
assumed to be in global gravitational collapse. A model combining the
calculation of small-scale collapse as in the above models with the
large-scale collapse of the whole cloud has been presented by
\cite[Zamora-Avil\'es et al. (2012, hereafter Z+12)] {ZA+12} and
\cite[Zamora-Avil\'es \& \VS\ (2014, hereafter ZV14)] {ZV14}. In what
follows we will refer to this as ``{\it the evolutionary model}''. In
it, in addition to the calculation of the instantaneous SFR as in the
models mentioned above, the evolution and collapse of the cloud (with
the associated systematic increase of its mean density) are
also followed in time. The cloud is assumed to be formed by the
collision of warm atomic streams, and to grow in mass by the continued
accretion of this material, which undergoes a phase transition to the
cold phase as it enters the cloud.  The cloud also develops turbulence
through a combination of the nonlinear thin-shell instability
(\cite[Vishniac 1994] {Vishniac94}) and the Kelvin-Helmholz,
Rayleigh-Taylor, and thermal instabilities (see, e.g., \cite[Heitsch \ea\
2005] {Heitsch+05}). Because the cloud is collapsing, the instantaneous
mass fraction at high densities increases in time, and therefore so does
the instantaneous SFR. Knowing the SFR at all times, the total mass in
stars can be computed and, assuming a standard IMF, the instantaneous
number of massive stars is also computed.  These feed back on the cloud
through ionizing UV radiation, and the rate of ionization can be
computed using standard prescriptions (\cite[Franco \ea\ 1994]
{Franco+94}), eventually destroying the cloud.

\begin{figure}[b]
\begin{center}
 \includegraphics[width=2.5in]{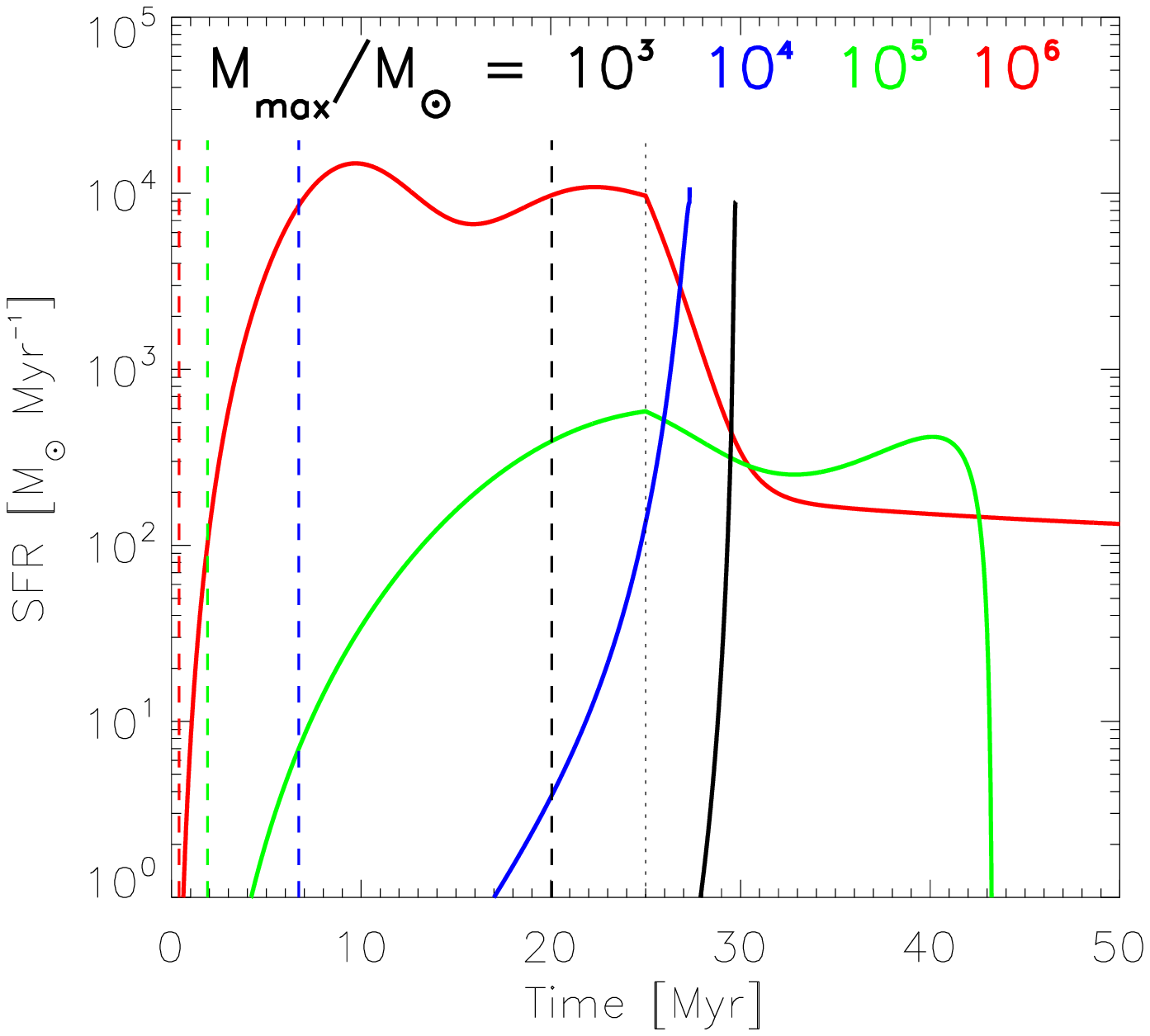} 
 \includegraphics[width=2.5in]{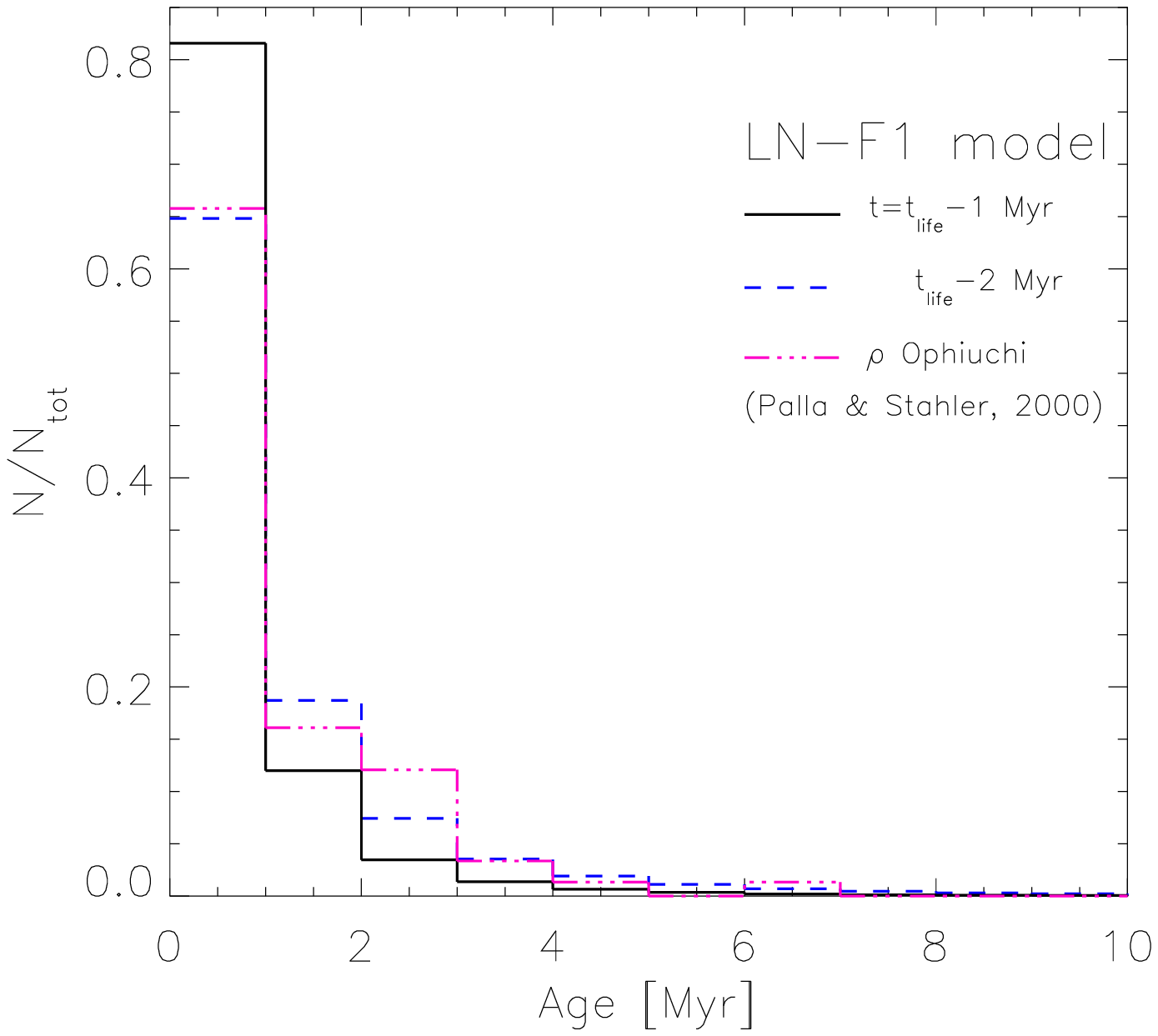} 
 \caption{{\it Left:} Time evolution of the SFR for clouds with
 $\Mmax=10^3$, $10^4$, $10^5$, and $10^6 \, \Msun$ (black, blue, green
 and red lines respectively). The vertical dashed lines represent the
 time at which the clouds start to form stars. The vertical black dotted
 line represents the time at which accretion onto the clouds from the
 warm medium stops (from ZV14). {\it Right:} Normalized stellar age
 histograms for a model cloud at 1 (solid black line) and 2 (dashed blue
 line) Myr before the cloud is destroyed by the ionizing stellar
 feedback. Also shown is the corresponding histogram for the embedded
 cluster in the $\rho$-Oph cloud (dash-dot-dot magenta line; \cite[Palla \&
 Stahler 2000] {PS00}) (from Z+12.).}
\label{fig:SFR}
\end{center}
\end{figure}

Contrary to the stationary equilibrium models, in the evolutionary model
the turbulence parameters are not free, but rather they are assumed to
have the (fixed) values typical of the cold atomic medium. Thus, the
turbulence is assumed to be {\it moderately} supersonic, with Mach
number $\Ms \sim 3$ (e.g., \cite[Heiles \& Troland 2003]{HT03}).  The
only remaining free parameter is the total mass involved in the
cloud-formation and evolution process ($\Mmax$). Figure \ref{fig:SFR}
shows, in the {\it left panel}, the evolution of the SFR for clouds of
various masses according to the model.
In the {\it right panel}, this figure
shows the histogram of stellar ages in a cluster that evolves according
to the evolution of the SFR predicted by the model in a 2000-$\Msun$
cloud, 1 and 2 Myr before the cloud is destroyed by the ionizing stellar
feedback. The figure also shows the normalized stellar age histogram in
the $\rho$-Ophiucus cloud (\cite[Palla \& Stahler 2000] {PS00}), showing
that the model at 2 Myr before destruction matches quite closely the
observed histogram. The match is expected to occur some time before
destruction because the clusters studied by \cite[Palla \& Stahler
(2000)] {PS00} are still embedded in their parent clouds, suggesting
that the destruction of the clouds is not completed yet.

\section{A simulation of cluster formation} \label{sec:sim}

The results from the model described in the previous section are
supplemented by those obtained in a numerical simulation of cloud
formation, evolution and destruction, presented by \cite[Col\'in
\ea\ (2013, hereafter C+13)] {Colin+13}. This simulation utilized the
N-body, AMR-Hydro code ART (\cite[Kravtsov et al.\ 1997] {KKK97}) with
(non-accreting) sink particles, adding a simplified treatment of radiative
transfer that allows a first-order estimate for each sink's Str\"omgren
radius. Cells within one Str\"omgren radius from a sink particle are
heated to $T=10^4$ K. Furthermore, a probabilistic SF
scheme was used, which allowed the sink particles to take a power-law
mass distribution, with a slope consistent with the Salpeter value. That
is, in this simulation, the sink particles correspond to individual
stars, with a realistic mass distribution. This allows following the
cluster dynamics realistically as well. The box size is 256 pc, and the
initial density and temperature in the box were $n = 1~\pcc$ and $T =
5000$ K, respectively. Two colliding streams of gas, of radius 64 pc and
length 112 pc and with opposite velocities of magnitude $5.9 \kms$ each,
were set to collide at the $x=0$ plane in the simulation. A turbulent
velocity field with a spectrum peaking at wavenumber $k = 8 \times 2
\pi/ \Lb$ (so that the fluctuation size scale is smaller than the radius
of the colliding streams) was applied to trigger instabilities in the
compressed layer. For more details, we refer the reader to C+13.

\begin{figure}[b]
\begin{center}
 \includegraphics[width=2.5in]{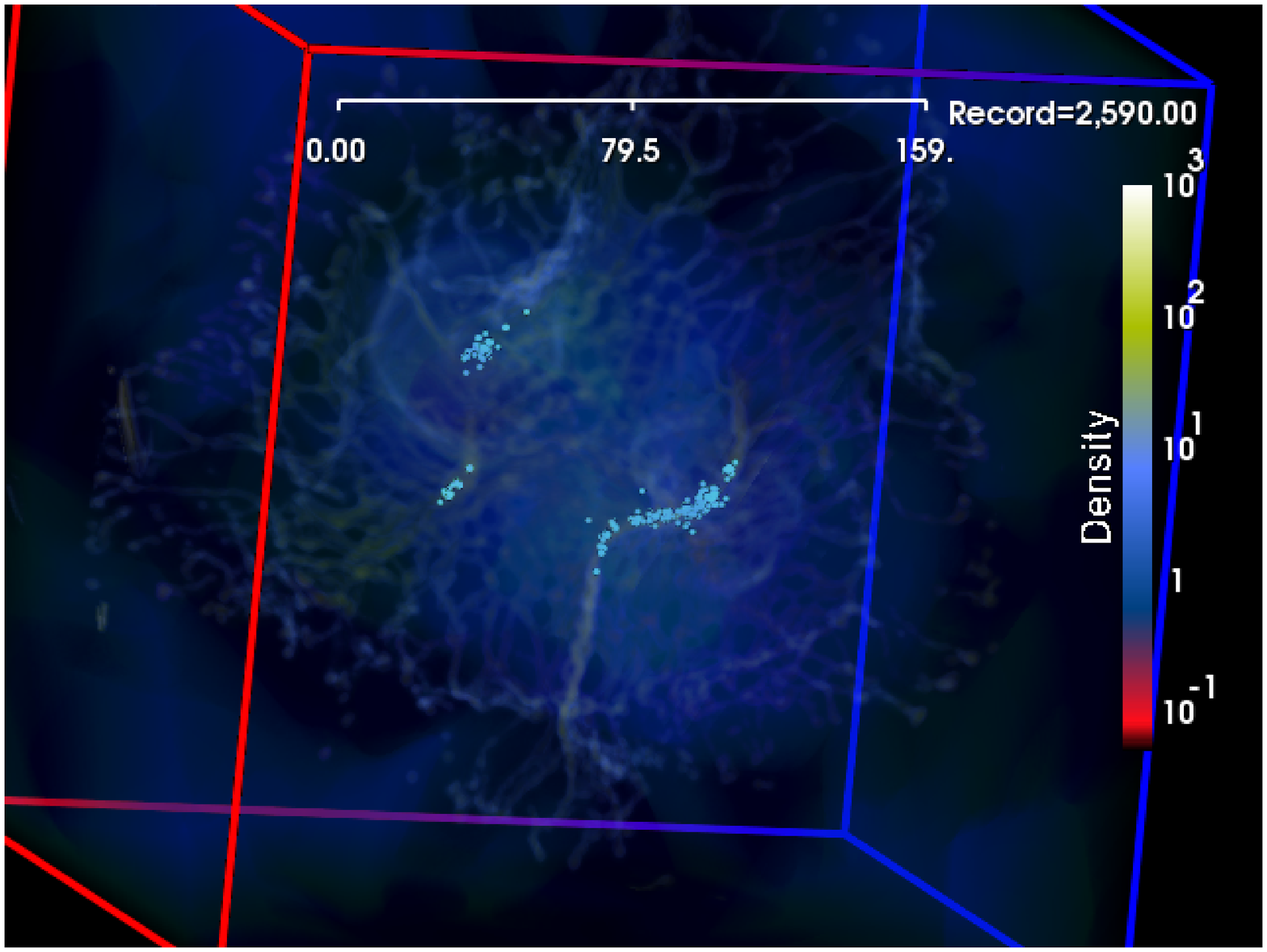} 
 \includegraphics[width=2.5in]{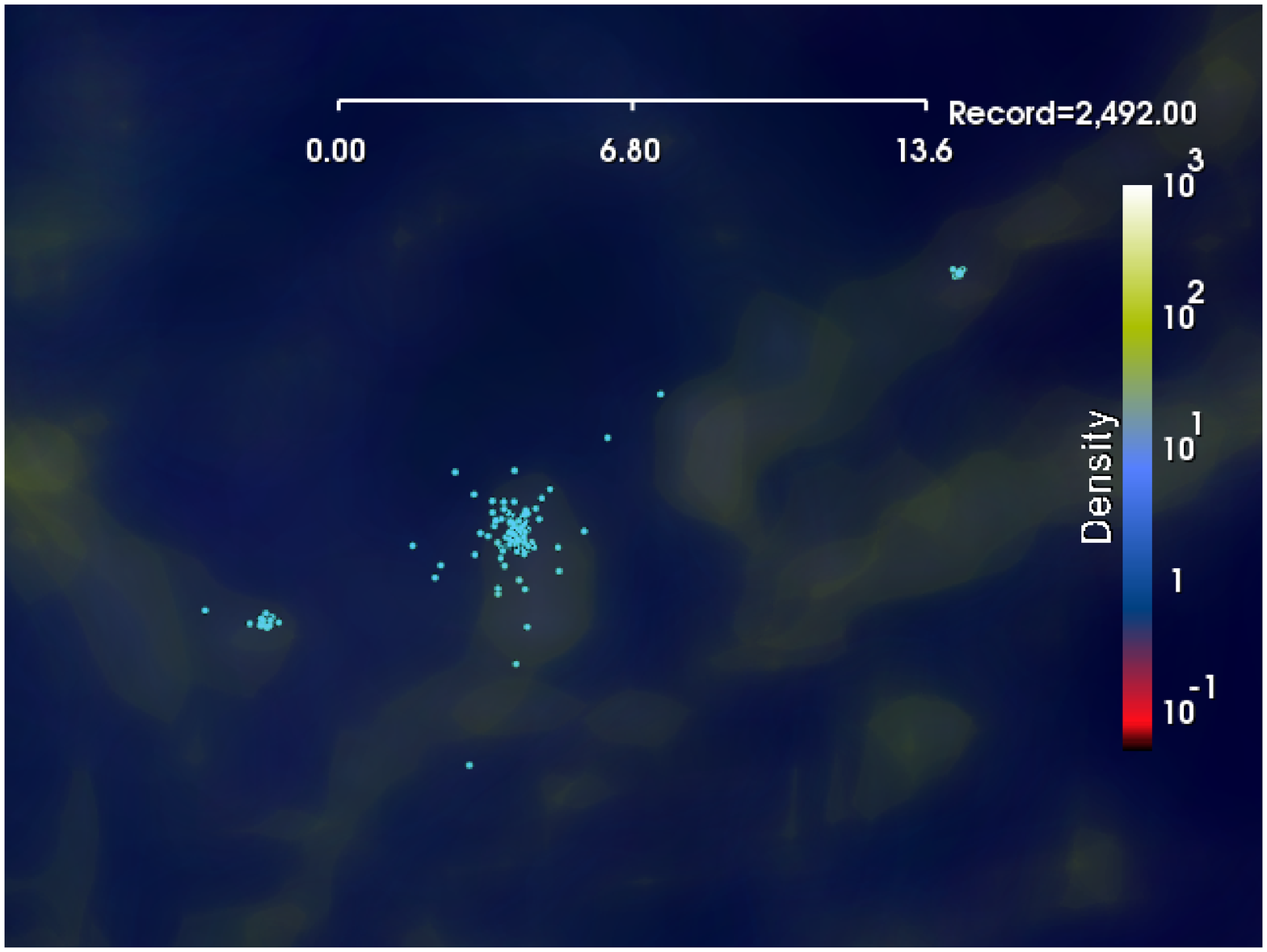} 
 \caption{Two views of the numerical simulation. {\it Left panel:} A
   global projected view at $t=25.9$~Myr, showing the three clusters
   that form within filamentary structures. The box size is 256 pc and
   the ruler shows a scale of 159 pc. The color bar indicates the local
   density in $\pcc$. {\it Right panel:} A zoom-in around the
   intermediate-mass cluster at $t=24.9$ Myr, showing that it is
   composed of three subunits at this time, which later ($t \sim 30$
   Myr) merge to form a single larger cluster. The central object in
   this image is the cluster whose evolution is shown in Fig.\
   \ref{fig:cluster_evol}, itself having formed from the merger of
   smaller subunits, and therefore illustrating the self-similar
   structure of the system. The ruler indicates a scale of 13.6 pc.}
 \label{fig:sim_imgs}
\end{center}
\end{figure}

In this simulation, three main clusters form, two of which are
sufficiently massive to disperse/evaporate the dense gas around them on
a timescale of $\sim 13$ Myr. Stars begin to form at $t \approx 18.9$
Myr, and the first \hii\ regions appear at $t \approx 24.2$ Myr. At $t
\approx 37.5$ Myr, the dense gas has been cleared from a radius of $\sim
70$ pc around the clusters. Figure \ref{fig:sim_imgs} shows, in the {\it
  left panel}, one snapshot of the simulation, at $t=25.9$ Myr into the
evolution, in which the growing \hii\ regions around the two most
massive clusters can be seen as faint shells around the clusters, while
the filamentary structure of the cloud is still noticeable in general.
The clusters themselves have formed along the main filamentary
structures in the cloud, and contain $10^3$--$10^4 \Msun$.

A crucial implication of the HGC scenario is that the filaments
constitute part of the large-scale gravitational collapse, funneling gas
into the cores within them, as observed by \cite[G\'omez \&
V\'azquez-Semadeni (2014)] {GV14} in a numerical simulation of cloud
formation and collapse. Moreover, these authors observed a hierarchy of
collapses within the filaments, so that {\it small clumps, which are
sometimes forming stars already, are falling onto larger-scale
ones}. This means that the late-stage clusters are assembled from
smaller sub-units that formed in the filaments in the periphery of the
trough of the large-scale gravitational potential well. This is
illustrated in the {\it right panel} of Fig.\ \ref{fig:sim_imgs}, which shows
a zoom-in around the second most massive cluster (hereafter, ``cluster
2'') of those seen in the left panel, at $t=24.9$ Myr. It can be seen
that this cluster is actually composed of three groups at this time,
which later (at $t \sim 30$ Myr) merge to form a single larger cluster. 

Operationally defining a group is a nontrivial task, because the gas
clumps continue to form stars over extended periods of time due to the
accretion from the filaments onto them. We thus define membership to a
group by a sequence of steps. At early times, when the groups are
clearly distinct by eye, we apply a ``friends-of-friends'' algorithm to
the stars to define the groups. Subsequently, as new stars are formed, we
assign them to the groups whose center of mass is closest to the new
star. Also, we define the radius of the group at each time as the
distance from its center of mass to the second most distant star.
Finally, we say that two groups have merged when the distance between
their centers of mass is smaller than the larger of the two radii. 

\begin{figure}[b]
\begin{center}
 \includegraphics[width=1.7in]{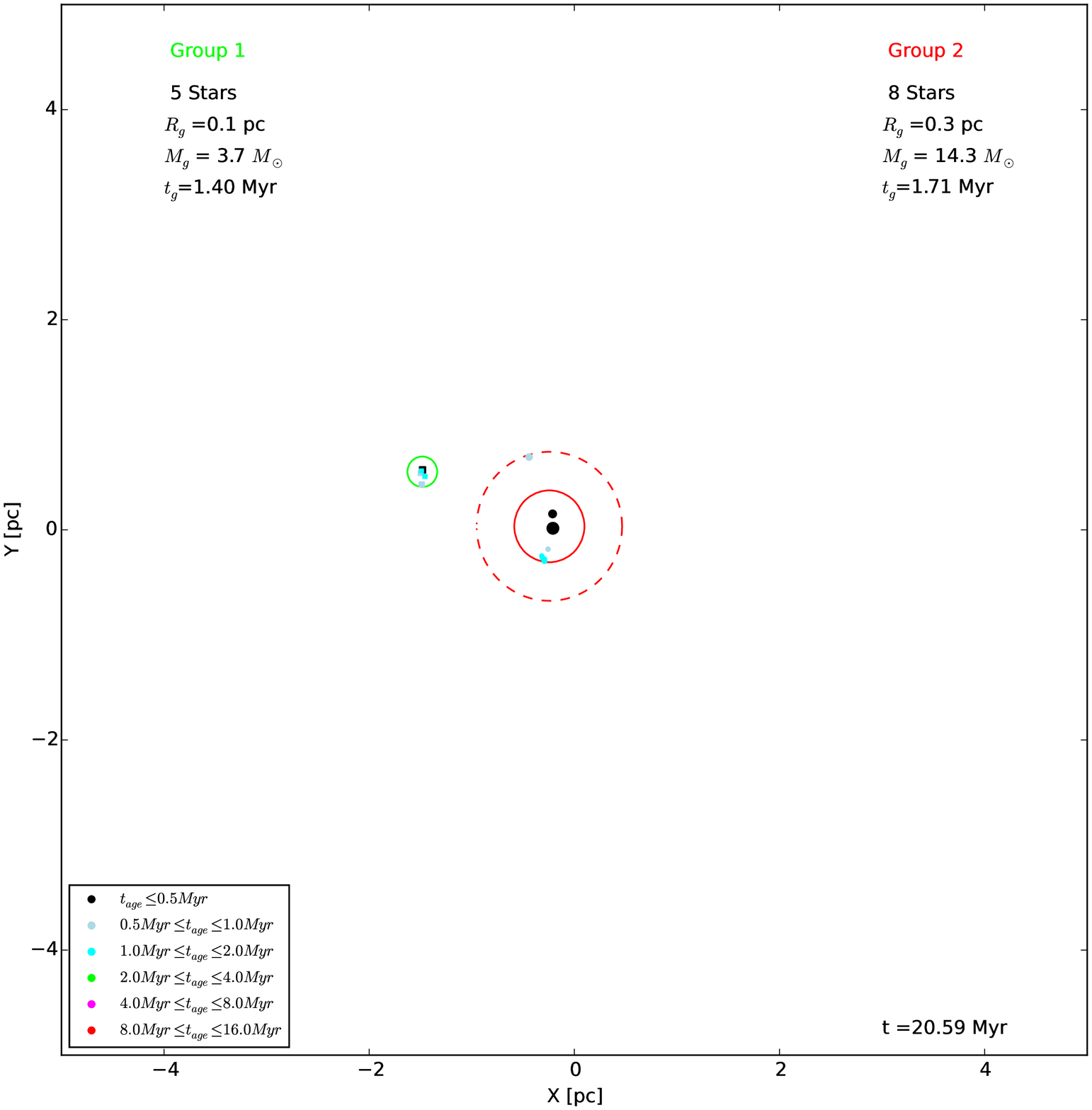} 
 \includegraphics[width=1.7in]{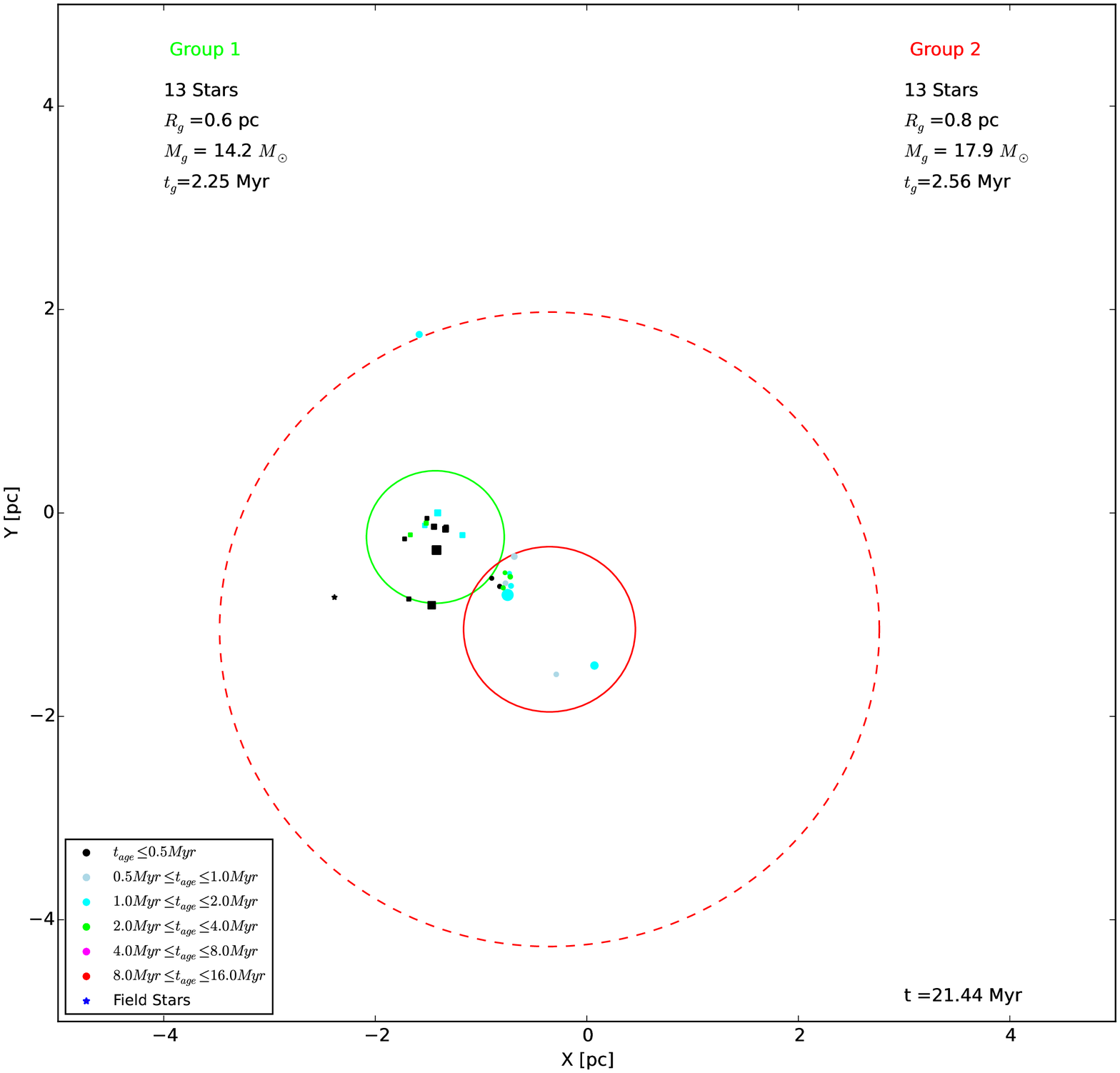} 
 \includegraphics[width=1.7in]{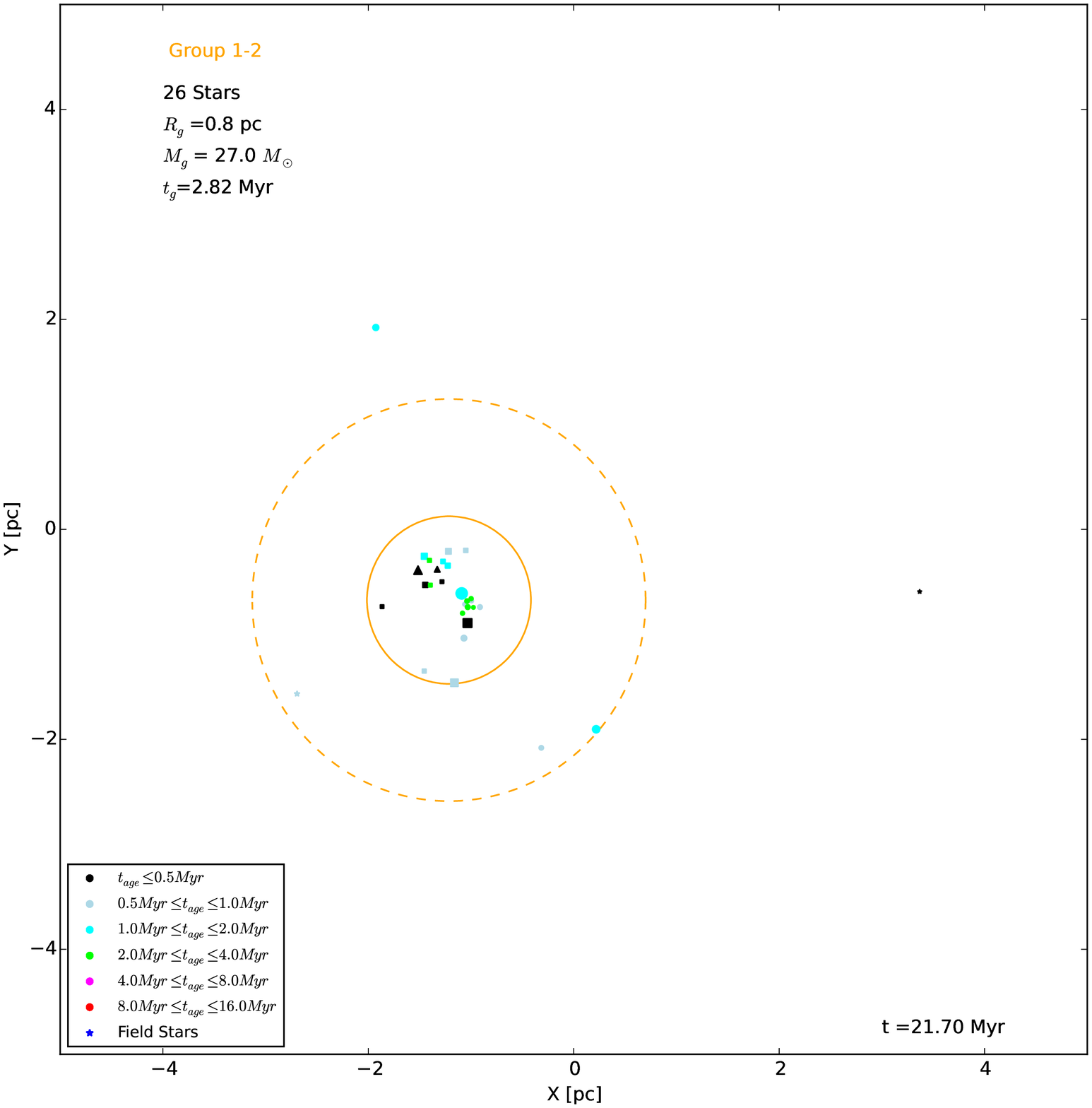} 
 \caption{Three snapshots of ``cluster 2'' at times $t=20.59$, 21.44,
   and 21.70 Myr in the simulation (respectively, left, middle, and
   right frames), showing the merging of two subunits to form the
   cluster.
 } 
 \label{fig:cluster_evol}
\end{center}
\end{figure}

The evolution and assembly of the central group in the {\it right
  panel} of Fig.\ \ref{fig:sim_imgs} is illustrated in Fig.\
\ref{fig:cluster_evol}, which shows the projected positions of the
sink particles on the $(x,y)$ plane at times $t=20.59$, 21.44, and
21.70 Myr into the evolution of the simulation in the {\it left, middle,} and
{\it right} panels, respectively. It is seen that, at the earlier time, the
system consists of two subgroups, which approach each other and finally
merge. In general, we observe this mechanism operating at all scales in our
simulation, with groups in turn consisting of subgroups, and so on,
implying that the clusters must exhibit a self-similar, fractal structure.

\section{Discussion} \label{sec:disc}

The results described above suggest that the multi-scale, hierarchical
gravitational collapse (HGC) of the clouds may be responsible for at
least three observed properties of open clusters: 1) that there
appears to be an increase of the SFR, so that the age distributions of
clusters generally have a tail of older stars (3-7 Myr), while most
stars are young (1-2 Myr; \cite[Palla \& Stahler 2000] {PS00}); 2) that
they appear to have a self-similar structure, observationally manifested
in an absence of a characteristic scale in the distribution of
protostellar separations (\cite[Bressert \ea\ 2010] {Bressert+10}). In
our simulation, this is manifested in the hierarchical membership of the
stellar units that conform a cluster; 3) that the clusters appear to
consist of subunits of slightly different ages that merge to form larger
structures (\cite [Kuhn \ea\ 2015b]{Kuhn+15b}; \cite[Rivera-G\'alvez et
al.\ 2015] {RG+15}). These are natural consequences of HGC in globally
collapsing clouds. The global collapse produces an increase of the SFR.
The hierarchical nature causes, at large scales, filamentary
structures that feed the star-forming clumps for extended periods of
time and, at small scales, it produces local star-forming sites that
fall into the troughs of the large-scale contracting structures.

\end{document}